\documentclass{elsart}
\usepackage{amsmath,epsfig}
\usepackage{doublespace}
\begin{document}
\begin{frontmatter}
\title{Perturbed soliton-like molecular excitations in a deformed DNA chain }
\author{V.~Vasumathi},
\author{M.~Daniel\corauthref{cor1}}
\ead{daniel@cnld.bdu.ac.in} 
\corauth[cor1]{Corresponding Author. Telephone:+91-431-2407057,  Fax:+91-431-2407093}
\address{ Centre for Nonlinear Dynamics, School of Physics, 
Bharathidasan University, Tiruchirappalli - 620 024, India.}
\date{} 
\begin{abstract}
     We study the  nonlinear dynamics of  a deformed Deoxyribonucleic acid (DNA) molecular chain which is governed by a perturbed sine-Gordon  equation coupled with a linear wave equation representing the lattice deformation. The DNA chain considered here is assumed to be deformed periodically which is the energetically favourable configuration, and the periodic deformation is due to the repulsive force between base pairs, stress in the helical backbones and due to the elastic strain force in both the strands. A multiple scale  soliton perturbation     analysis  is carried out to solve the perturbed sine-Gordon equation and the resultant perturbed kink and antikink solitons represent  open state configuration with small fluctuation.  The perturbation due to periodic deformation of  the lattice changes the velocity of the soliton. However,  the width of the soliton remains unchanged.
\end{abstract}
\end{frontmatter}
\maketitle
\section{Introduction}
 Deoxyribonucleic acid (DNA) plays an important role in the
 conservation and transformation of genetic information in 
  biological systems \cite{ref1}. Opening of base pairs in DNA double helix is related to functions like transcription and replication. Base pair opening via nonlinear molecular excitations has been understood by several authors \cite{ref2,ref11,ref12,ref13,ref14,ref3,ref4,refpb1,ref14a} by proposing different models.    Among them the models proposed and used by
    Yomosa \cite{ref11,ref12}, as well as by Takeno and Homma \cite{ref13,ref14} were based on rotation of 
  bases in a plane normal to the helical axis of DNA, and the nonlinear molecular excitations were governed by kink-antikink solitons.  Following Takeno and Homma, recently  several authors \cite{refsa,g1,g2,ref36a,g3,g4,ref18,ref26} studied  soliton-like molecular excitations in DNA by taking into account the rotation of bases.  In all the above studies, both the strands of the DNA double helix were considered as  rigid lattices.  However, in nature the force between purine bases in consecutive base pairs is 
	 repulsive,  and this force is resisted by stress in the helical 
	 backbones of DNA and also, the main-chain torsion angle indicates that 
	 there are elastic strain forces in both the strands \cite{ref15,ref16}. The dynamics of this non-rigidity of the strands gives rise to phonons which  also play an important role in energy transfer in biological systems. In a different context, Davydov \cite{ref16a,refdvs1} proposed a model for energy transfer in alpha helix protein molecules and he found that the  propagation of molecular vibrations induce longitudinal sound waves (phonons), which in turn provides a potential well that prevents vibrational dispersion, and this coupled excitation propagates as a soliton without loss of energy along hydrogen bonding spines of the alpha helical protein. Thus, the study of nonlinear molecular excitations in DNA double helical chains coupled with phonons in strands or in other words the influence of
	   non-rigidity of the strands in molecular  excitations has  become an 
	   important task which requires a detailed investigation.  In this direction, recently, Xiao and his co-workers \cite{ref17} studied the influence of longitudinal vibration on the soliton excitations in DNA double helix by considering the dynamic plane base rotator model of Takeno-Homma \cite{ref13,ref14}, and  by including the  longitudinal vibration  and its coupling with hydrogen bonds and stacking. It was  shown that the dynamics in this case is governed by a perturbed sine-Gordon  equation in the continuum limit,  which upon solving using the method of successive approximation by iterations gives soliton under first order approximation,  which shows that   the effect of longitudinal vibration of the lattice on soliton is small. However, they failed to find  the variation of the soliton parameters such as velocity and width explicitly during propagation under iterations. 
	    Therefore, in the present paper, we study the nonlinear molecular excitations in  DNA  double helix with non-rigid elastic strands, by  solving the dynamical equation  using  direct soliton perturbation theory, which provides the variation of velocity and width of the soliton under perturbation in explicit analytical terms.  The paper is organised as follows. In section 2, we  consider the model Hamiltonian for the above DNA double helix  and derive the dynamical equations. In the continuum limit,  the dynamical equations  reduce to a perturbed sine-Gordon equation coupled with a linear wave equation representing the longitudinal lattice vibration and this is treated in section 3. In section 4, 
	  a multiple scale soliton perturbation theory is developed to investigate the effect of lattice deformation on the open state configuration of DNA represented 
	 in terms of  kink-antikink solitons of the sine-Gordon equation.  The results are concluded in section 5.

\section{Model and  dynamical equations}
We consider the B-form of a DNA double helix with flexible strands,  and investigate the nonlinear molecular excitations by considering a plane-base rotator model. In Fig. (1a) we   have presented a sketch of the DNA double helix with z-axis parallel to the helical axis. In the figure, $S$ and $S'$  represent the two complementary strands in the DNA double helix, and each arrow in the figure represents the direction of the bases attached
     to the strand and the dots between arrows represent  the net hydrogen bonding effect between the 
     complementary bases.     
       In Fig. 1(b), we present a horizontal projection of the $n^{th}$ base pair in the xy-plane in which
      $Q_n $ and $Q'_n$ denote the  tips of the $n^{th}$ bases, and $P_n$ and $P'_{n}$ represent the points where the $n^{th}$ bases are attached to the strands $S$ and
     $S'$ respectively.

  The  conformation and stability of  DNA double helical molecular chains are mainly determined by the  stacking  between  the adjacent bases in the strands and the hydrogen bonds between the complementary bases. 
The  Hamiltonian under plane base rotator model involving stacking and hydrogen bonds in terms of the rotational angles $\phi_n$ and $\phi'_n$ of  the $n^{th}$ bases (see Fig. 1(b)) in the case of rigid strands as proposed by  Yomosa \cite{ref11,ref12} and further developed by Takeno and Homma  \cite{ref13,ref14} is written as
\begin{eqnarray}
  H_r&=&\sum_n\left[ \frac{I}{2}( {\dot\phi_n}^{2}+ {\dot\phi_n}^{'2}) +J[2-\cos (\phi_{n+1}-\phi_n)-\cos(\phi'_{n+1}-\phi'_n)]\right.\nonumber\\
 &&\left.-\alpha[1-\cos(\phi_n-\phi'_n)]\right],\label{eq1}  
 \end{eqnarray}
where $I$ is the moment of inertia of the nucleotides around the axes at $P_n$ and $P'_n$  in the strands $S$ and $S'$ respectively and thus, the first two terms represent the kinetic energies 
  of the rotational motion  of the $n^{th}$  nucleotide bases.  In Eq. (\ref{eq1}) overdot represents time derivative.
Further, the terms proportional to $J$   in Hamiltonian (\ref{eq1})  represent the  stacking energy between the $n^{th}$ base 
and its nearest neighbours  in the strands $S$ and $S'$  and $\alpha$  represents a measure of the
interstrand interaction or hydrogen bonding energy between the complementary bases respectively.  Pople's formula in which the mean energy of the distorted hydrogen bonds is approximately represented in the above form \cite{pople}.
   However, in nature DNA strands are not rigid but flexible and hence, we assume that the two strands  deform elastically and the resultant phonons  couple to the stacking and hydrogen bonds.  Hence, the part of the Hamiltonian  corresponding to the phonon energy,  and the energy due to its coupling with the stacking and hydrogen bonds is written as
\begin{subequations}
\begin{eqnarray}
  H_{ph}&=&\sum_n\left[ \frac{p_n^{2}}{2M}+\frac{{p'}_{n}^{2}}{2M}+K[(y_{n+1}-y_n)^{2}+(y'_{n+1}-y'_n)^{2}]\right],\label{eq2a} \\
H_{r-ph}&=&\sum_n\left[\beta\{ (y_{n+1}-y_n)[1-\cos (\phi_{n+1}-\phi_n)]+ (y'_{n+1}-y'_n)\right.\nonumber\\
 &&\left.\times[1-\cos(\phi'_{n+1}-\phi'_n)]\}-\gamma(y_{n+1}-y_{n-1})[1-\cos(\phi_n-\phi'_n)]\right],\label{eq2b}  
 \end{eqnarray} 
\end{subequations}
  where $p_n=M\dot y_n$ and $p'_n=M\dot y'_n$. In Eqs. (2),  $y_n $ and $y'_n$ represent the longitudinal displacements of the $n^{th}$ nucleotides from the equilibrium position in the two strands,  and $M$ is the uniform mass of the nucleotide. $K$ is the longitudinal elastic constant along the double helical main chain. The stacking energy  depends on the distance between the $n^{th}$ and $(n+1)^{th}$ base,   and  the strength of the hydrogen bonds   depends symmetrically on the distance between the $(n-1)^{th}$ and $(n+1)^{th}$ bases. Thus,  $\beta$ and $\gamma$ measure the coupling strengths between phonon and stacking as well as hydrogen bonds respectively. The interaction Hamiltonian $H_{r-ph}$ in Eq.(\ref{eq2b}) is chosen to represent the change in stacking energy and hydrogen bonds energy caused by the change in the displacement of the nucleotides along the two strands. As we are going to study the dynamics in the low temperature and long wavelength limit, it is appropriate to consider linear coupling of phonon to the stacking and hydrogen bonds.
      Now, using the Hamiltonians (\ref{eq1}), (\ref{eq2a}) and (\ref{eq2b}), the total Hamiltonian $H$ for the system is written as 
$H=H_r+H_{ph}+H_{r-ph}$ and the corresponding  Hamilton's equations of motion  take the form
  \begin{subequations}
  \begin{eqnarray}
  I\ddot\phi_n&=&[J+\beta(y_{n+1}-y_n)] \sin (\phi_{n+1}-\phi_n)-[J+\beta(y_{n}-y_{n-1})]\nonumber\\
&&\times\sin
  (\phi_n-\phi_{n-1})+[\alpha +\gamma(y_{n+1}-y_{n-1})]\sin (\phi_n-\phi'_n),\label{eq3a}\\  
  I\ddot\phi'_n&=&[J+\beta(y_{n+1}-y_n)] \sin (\phi'_{n+1}-\phi'_n)-[J+\beta(y_{n}-y_{n-1})]\nonumber\\
&&\times\sin(\phi'_n-\phi'_{n-1})+[\alpha +\gamma(y_{n+1}-y_{n-1})]\sin (\phi'_n-\phi_n),\label{eq3b}\\
M\ddot y_n&=&2K(y_{n+1}-2y_n+y_{n-1})-\beta[\cos(\phi_{n+1}-\phi_n)-\cos(\phi_n-\phi_{n-1})]\nonumber\\
  &&+\gamma[\cos(\phi_{n+1}-\phi'_{n+1})-\cos(\phi_{n-1}-\phi'_{n-1})],\label{eq3c}\\
M\ddot y'_n&=&2K(y'_{n+1}-2y'_n+y'_{n-1})-\beta[\cos(\phi'_{n+1}-\phi'_n)-\cos(\phi'_n-\phi'_{n-1})]\nonumber\\
  &&+\gamma[\cos(\phi_{n+1}-\phi'_{n+1})-\cos(\phi_{n-1}-\phi'_{n-1})].\label{eq3d}
\end{eqnarray}
  \end{subequations}
  Eqs. (\ref{eq3a}-\ref{eq3d}) describe the dynamics of DNA with deformable strands at the discrete level
by  considering the dominant angular rotation of  bases  in a plane  normal to the helical axis of the DNA,  and ignoring all other small motions of the bases combined with longitudinal motion of the nucleotides.\\
 \section{Soliton and base pair opening} 
     It is expected that   the difference  in the  angular rotation of  bases with respect to neighbouring bases along the two strands in DNA  namely $ (\phi_{n\pm 1}-\phi_n)$ and $(\phi'_{n\pm 1}-\phi'_n)$ are  small \cite{ref13,ref14}. Also, as the length of the DNA chain is very large due to the presence of large number of bases compared to the distance between the neighbouring base pairs, we  make a continuum approximation   by  introducing two fields of rotational angles  $\phi_n(t)\rightarrow\phi(z,t),~ \phi'_n(t)\rightarrow\phi'(z,t)$ and two fields of longitudinal displacement  $ y_n(t)\rightarrow y(z,t)$ and $ y'_n(t)\rightarrow y'(z,t)$ along the strands  where $z=na$ with $l$, the lattice parameter. Also, we make the expansions for $
 \phi_{n\pm1}=\phi(z,t)\pm a\frac{ \partial{\phi}}{\partial{z}}+\frac{
 a^{2}}{2!} \frac{\partial^{2}{\phi}}{{\partial{z}}^{2}}\pm\frac{
 a^{3}}{3!} \frac{\partial^{3}{\phi}}{{\partial{z}}^{3}}+..., $
and similar expansions  for $\phi'_{n\pm 1},~y_{n\pm 1}$ and $
 y'_{n\pm 1}$. Thus, in the   continuum limit under small angular rotation of bases Eqs. (3)  upto O($a^3$)   become
\begin{subequations}
\begin{eqnarray}
\phi_{\hat t\hat t}&=&\phi_{zz}-\frac{1}{2} \sin (\phi-\phi')+\epsilon [\beta(y_z \phi_{z})_{z}
+\frac{\hat\gamma}{2} y_z\sin (\phi-\phi')]
,\label{eq4a}\\
\phi'_{\hat t\hat t}&=&\phi'_{zz}-\frac{1}{2} \sin(\phi'-\phi)+\epsilon [\beta(y'_z \phi'_{z})_{z}+\frac{\hat\gamma}{2} y'_z\sin(\phi'-\phi)]
,\label{eq4b}\\
y_{\hat t\hat t}&=&v^2 y_{zz},\label{eq4c}\\
y'_{\hat t\hat t}&=&v^2 y'_{zz},\label{eq4d}
\end{eqnarray} 
\end{subequations}
where $\epsilon =\frac{a}{J},~v^{2}=\frac{2KI}{JM}$ and the suffices $\hat t$ and $z$ in Eqs.(4) represent  partial time and spatial derivatives and the rescaled $a$ is dimensionless. While writing the above equations we have chosen $\alpha=-\frac{1}{2}Ja^{2}$ and also, the parameter $\gamma$  is rescaled as $\hat\gamma= \frac{a^2\gamma}{4}$. Further, before writing Eqs. (4a-d), we have divided the full equations by $J a^2$ and rescaled the time variable as  
  $\hat{t}=\sqrt{\frac{Ja^{2}}{I}} t$. It is more convenient to describe the transverse motion of the bases in  DNA strands in terms of the center of mass co-ordinates. For this, we rewrite Eqs. (4)  by subtracting and adding  the first two and the last two equations  respectively.  Further, to commence the open state configuration of DNA, the two complementary bases are expected to rotate in opposite  directions  and both the strands are assumed to vibrate in the same direction so   that $\phi'=-\phi$  and  $y'=y$. Under these conditions,  we obtain
\begin{subequations}
\begin{eqnarray}
\Psi_{\hat t\hat t}-\Psi_{zz}+\sin\Psi=\epsilon [\beta(y_z \Psi_{z})_{z}+\hat\gamma y_z\sin\Psi],\label{eq5a}\\
y_{\hat t\hat t}-v^2 y_{zz}=0,\label{eq5b}
\end{eqnarray}
\end{subequations}
   where $\Psi=2\phi$.
    Eqs.~(5) describe the dynamics of bases under  a plane-base  rotator model of  DNA double helical chain  with the   deformed  strands. The terms proportional to $\beta$ and $\hat\gamma$ in the right hand side of Eq. (\ref{eq5a}) represent the coupling of phonon to the stacking and hydrogen bonds respectively.\\
 When  $\epsilon=0$, Eqs.~(\ref{eq5a}) and (\ref{eq5b})  are decoupled,  and Eq. (\ref{eq5a}) reduces to the completely integrable sine-Gordon equation which admits kink and antikink-type of soliton solutions,   and hence we call Eq. (\ref{eq5a}) in its present form  as a perturbed sine-Gordon equation.
      The integrable sine-Gordon equation $(\epsilon=0)$ was originally solved for N-soliton
      solutions using the most celebrated Inverse Scattering Transform (IST) method by Ablowitz and his co-workers \cite{ref19}.    
       The kink and antikink  one soliton solution of the integrable sine-Gordon equation
(Eq.(\ref{eq5a}) when $\epsilon=0$) can be written as
 \begin{eqnarray}
  \Psi(z,\hat t)=4 \mbox{arc}\tan{\exp[\pm m(z-v\hat t)]},~~ m^{-1}= \sqrt{1-v^{2}}. \label{eq11}    
 \end{eqnarray} 
In Eq.(\ref{eq11}), while the upper sign corresponds to kink soliton, the lower sign represents the antikink
soliton.
Here,  $v$ and $m^{-1}$ are real parameters that determine the  velocity and width of the soliton 
respectively.
 The kink and antikink  one soliton solutions  as given above  are depicted in Figs. 2(a) and 2(b).
       The kink-antikink soliton of the sine-Gordon equation describes an open
       state in  DNA double helix which is schematically represented in Fig. 2(c). In this figure the base
        pairs are found to open locally in the form of kink-antikink 
	shape in each strand and  the opening is found to propagate along the direction of the helical
	 axis.

Eq.(\ref{eq5b}) is the well known one-dimensional linear wave equation which admits wave solution in the form $
y=f(z-v \hat t)+g(z+v \hat t),$ where $f$ and $g$ are arbitrary functions. Now, the problem boils down to solving the perturbed sine-Gordon equation (\ref{eq5a}) after using   the wave solution `$y$'  obtained by solving  Eq. (\ref{eq5b}).

\section{Effect of  elastic deformation of strands on base pair opening}  
\subsection{A perturbation approach}   
When the phonon due to elastic deformation of the strands is coupled to the DNA molecular excitations, it is expected to perturb the kink and antikink solitons  in DNA which correspond 
	 to the open state configuration. It is further expected that the perturbation due to phonon coupling modifies the shape, width and velocity of the soliton as it propagates along the helical chain. In  order to understand this, we solve Eq. (\ref{eq5a}) using a suitable perturbation method. One of the most powerful techniques in dealing with perturbed soliton
	 is the soliton perturbation theory  which is based on the IST method.
	 However, as the method is
	very sophisticated it is very difficult to use the same in
	several cases. In view of this, many  authors
	used different types of direct  methods to study  soliton perturbation (see for e.g. refs.
	\cite{ref20,ref21,ref22,ref23,ref24}).
		  In the present paper, we use one such direct perturbation method 
	   to solve the perturbed sine-Gordon equation (\ref{eq5a}) to understand
	    the effect of phonon interaction
	  on the open state  configuration of DNA,  which is also dealt in reference 
	 \cite{ref24} in a different context,  and also by the present authors recently while studying the nonlinear
molecular excitations in an inhomogeneous DNA \cite{ref18}. 
	 The procedure we adapt here is based on the derivative
	expansion method to linearize the perturbed sine-Gordon equation in the
	coordinate frame attached to the moving frame. The parameters of the
	kink-antikink soliton are assumed to depend on a slow time scale in order to
	eliminate the secular terms. The linearized equations will be solved using
	the method of separation of variables  which  will be ultimately related to a generalized eigenvalue problem, the eigenfunctions of which form the bases of the
	perturbed solution.  In  the following we  use the above approach to find
	the perturbed  soliton solution of Eq. (\ref{eq5a}). 
\subsection{Linearization of  the perturbed sine-Gordon equation}
  In order to study the effect of perturbation due to phonon interaction on the soliton,  the time variable $\hat t$ is transformed
into several variables as $t_n=\epsilon^{n} \hat t$,  where n=0, 1, 2,... and $\epsilon$ is a very small parameter.
 In view of this, the
time derivative  and $\Psi$ in Eq.~(\ref{eq5a}) are replaced by the expansions
$
\frac{\partial }{\partial \hat t}=\frac{\partial }{\partial t_0}+\epsilon~
\frac{\partial }{\partial t_1}+\epsilon^2 \frac{\partial }{\partial
t_2}+...$ and $\Psi=\Psi^{(0)}+\epsilon \Psi^{(1)}+\epsilon^2 \Psi^{(2)}+...$
and  we equate
the coefficients of different powers of $\epsilon$. Thus at $O(\epsilon^{(0)})$ we obtain 
\begin{eqnarray}
 \Psi^{(0)}_{t_0t_0}-\Psi^{(0)}_{zz}+\sin\Psi^{(0)}=0, \label{eq12}
\end{eqnarray}
for which the one soliton  solution takes the form     $\Psi^{(0)}(z,t_0)= 4 \mbox{arc}\tan\exp\zeta,~\zeta= \pm m_{0}(z-\xi),~\xi_{t_0}= v_{0},$
where $v_{0}$ is the velocity of the soliton in the $t_{0}$ time scale. Due to perturbation, the soliton parameters  namely $m$ and $\xi$ are now treated
as functions of the slow time variables $t_0,t_1,t_2, .... $ However, $m$ is treated as independent
of $t_0$. 
 The equation at   $ O(\epsilon^{(1)})$  is of the form
\begin{eqnarray}
 \Psi^{(1)}_{\tau\zeta}-\Psi^{(1)}_{\zeta\zeta}+(1-2\mbox{sech}^{2}\zeta)\Psi^{(1)}=F^{(1)}
 (\zeta,\tau),\label{eq13}
 \end{eqnarray}
 where
\begin{eqnarray}
  F^{(1)}&=& 2\beta\left[ y_{\zeta} \mbox{sech}\zeta \right]_{\zeta}+2~b \hat\gamma y_{\zeta}\tanh\zeta\mbox{sech}\zeta \nonumber\\
&&+ 4v_{0}  \left[m_{t_{1}}+(m^{2}\xi_{t_{1}}-\zeta m_{t_{1}})\tanh \zeta \right]\mbox{sech}\zeta
 .\label{eq14}
\end{eqnarray} 
 While writing the above equation  we have  replaced $\sin\Psi^{(0)}$ by $2b\tanh\zeta$ $\mbox{sech}\zeta$, where $b=\pm 1$, which can be derived using the unperturbed solution given below Eq. (\ref{eq12}),  and we  have also used the transformation
  $\hat \zeta=m(z-v t_{0})$ and $ \hat t_{0}=t_{0}$ to represent everything in a co-ordinate
  system that is moving with the soliton. Further, we have used  another set of
  transformations given by $\tau=\frac{\hat t_0}{2m}-\frac{(1+v)\hat \zeta}{2}$ and $\zeta=\hat \zeta$ for our later convenience.\\
 The solution of Eq.~(\ref{eq13}) is searched by assuming 
$\Psi^{(1)} (\zeta,\tau)=X(\zeta)T(\tau)$ and $
F^{(1)} (\zeta,\tau)= X_{\zeta}(\zeta)H(\tau).$
Substituting the above in Eq.~(\ref{eq13}),   we obtain
\begin{eqnarray}
\frac{1}{X_{\zeta}}[X_{\zeta\zeta}+(2 \mbox{sech}^{2}\zeta-1)X]=\frac{1}{T}[T_{\tau}-H(\tau)] \label{eq14a}
\end{eqnarray} 
In Eq. (\ref{eq14a}), while the left hand side is independent of $\tau$, the right hand side is independent of the variable $\zeta$. Therefore, we can equate the left and right hand sides of Eq. (\ref{eq14a}) to a constant, say $\lambda_0$ and write
\begin{eqnarray}
X_{\zeta\zeta}+(2 \mbox{sech}^{2}\zeta-1)X =\lambda_{0} X_{\zeta},\quad
T_{\tau}-\lambda_{0} T = H(\tau).\label{eq15}
\end{eqnarray}
  Thus, the problem of constructing the
perturbed soliton at this moment turns out to be solving 
 Eq. (\ref{eq15}) by constructing the eigenfunctions and 
finding the eigenvalues.
 The first part of Eq. (\ref{eq15}) is  a generalized eigenvalue problem,  which is not a self-adjoint eigenvalue problem and differs from the normal eigenvalue problem, with $X_{\zeta}$ in the right hand side instead of $X$. For solving the eigenvalue problem, we consider it in a more general form by writing
 \begin{eqnarray}
 L_{1} X=\lambda \tilde{X} ,~ L_{1}=\partial_{\zeta\zeta}+ 2\mbox{sech}^{2}\zeta -1, \label{eq17}
\end{eqnarray}
 where $\lambda$ is the eigenvalue. In order to solve Eq.(\ref{eq17}) 
 for $X$, we  also consider the following   eigenvalue problem.
\begin{eqnarray}
\qquad L_{2} \tilde{X}=\lambda X, \label{eq18}
\end{eqnarray}
 where $L_{2}$ is to be determined. Now, by combining the  above two eigenvalue problems we get
$L_2 L_1 X= \lambda^2 X,~~L_1 L_2 \tilde X= \lambda^2 \tilde X$. From these  expressions we conclude that $L_1 L_2$ is the adjoint of $L_2 L_1$ and also $X$ and $\tilde X$ are expected to be adjoint eigenfunctions. Hence, by solving the coupled eigenvalue problem we can find the eigenfunction $X$. Here $L_1$ is known  and is given in Eq. (\ref{eq17}), but the operator $L_2$ is still unknown. So, by experience we choose $L_{2}=\partial_{\zeta\zeta}+6\mbox{sech}^{2}\zeta -1$.\\ 
Now, in order to find  the eigenfunctions by solving  Eqs. (\ref{eq17}) and (\ref{eq18}) we choose the eigenfunctions as
\begin{eqnarray}
X(\zeta,k)=p(\zeta,k) e^{ik\zeta},~~
\tilde {X}(\zeta,k)=q(\zeta,k) e^{ik\zeta}, \label{eq19}
\end{eqnarray}
where $k$ is the propagation constant. On substituting the above  in  Eqs. (\ref{eq17}) and (\ref{eq18}) in the asymptotic limit,  we obtain the eigenvalue as $\lambda=-(1+k^{2})$.  In order to find  the eigenfunctions,  we expand  $p(\zeta,k)$ and $q(\zeta,k)$ as
\begin{subequations}
\begin{eqnarray}
p (\zeta,k)&=&p_0+p_1 \frac{\sinh\zeta}{\cosh\zeta}+p_2 \frac{1}{\cosh^{2}\zeta}
+p_3
\frac{\sinh\zeta}{\cosh^{3}\zeta} 
+p_4\frac{1}{\cosh^{4}\zeta}+...,\label{eq20a}\\
q (\zeta,k)&=&q_0+q_1\frac{\sinh\zeta}{\cosh\zeta}+q_2 \frac{1}{\cosh^{2}\zeta}
+q_3
\frac{\sinh\zeta}{\cosh^{3}\zeta} 
+q_4\frac{1}{\cosh^{4}\zeta}+...,\label{eq20b}
\end{eqnarray}
\end{subequations}
where $p_j$ and $q_{j}$, j=0,1,2,... are functions of $k$ to be determined.  Substituting Eqs. (\ref{eq19}), (\ref{eq20a}) and (\ref{eq20b}) in Eqs. (\ref{eq17}) and (\ref{eq18}) and collecting the coefficients of $ 1,~
\frac{\sinh\zeta}{\cosh\zeta},~ \frac{1}{\cosh^{2}\zeta}$,...   we get a set of simultaneous equations. On solving those equations by assuming $p_j=q_j=0$ for $j\geq 3$,  we obtain the   eigenfunctions as
\begin{subequations}
\begin{eqnarray}
X(\zeta,k)&=&\frac{(1-k^{2}-2ik\tanh\zeta)}{\sqrt{2\pi}(1+k^{2})}e^{ik\zeta},
\label{eq21a}\\
\tilde{X} (\zeta,k)&=&
\frac{(1-k^{2}-2ik\tanh\zeta-2\mbox{sech}^{2}\zeta)}{\sqrt{2\pi}(1+k^{2})} e^{ik\zeta}.
 \label{eq21b}
\end{eqnarray}
\end{subequations}
On comparing Eqs. (\ref{eq21a}) and (\ref{eq21b}) we can write $\tilde{X}(\zeta,k)=\frac{X(\zeta,k)}{ik}$. Now, using this in the right hand side of Eq. (\ref{eq17}) and comparing the resultant equation with Eq. (\ref{eq15}), we obtain $\lambda_0=\frac{i(1+k^2)}{k}$.\\
The second part of Eq. (\ref{eq15}) is a linear inhomogeneous differential equation and it can be solved using known procedures \cite{ref18}. The solution reads
\begin{eqnarray}
  T(\tau,k)=\frac{1}{i\lambda_{0} k(1+k^{2})}\int_{-\infty}^{\infty}
d\zeta  F^{(1)} (\zeta,\tau){X}^{\ast}(\zeta,k)
(e^{\lambda_{0}
[\tau+\frac{(1+v)}{2}\zeta]}-1),\label{eq16}
\end{eqnarray}
 The first order correction to the soliton can be computed using the following expression.
 \begin{eqnarray}
\Psi^{(1)}(\zeta,\tau)=\int_{-\infty}^{\infty} X(\zeta,k) T(\tau,k)
dk+\sum_{j=0,1}X_{j} (\zeta) T_{j}(\tau).\label{eq22}
\end{eqnarray}
Here $X(\zeta,k)$ and $ T(\tau,k)$ are known continuous eigenfunctions which are given in Eqs. (\ref{eq21a}) and (\ref{eq16}). However, the discrete eigenstates  $X_{0}, X_{1}$ and $T_{0},T_{1}$ are unknown. $X_{0} $ and $ X_{1}$  are the two   discrete eigenstates for the discrete eigenvalue $\lambda=0$ and these states  can be found out  using the completeness  of the continuous eigenfunctions as
\begin{eqnarray}
 X_{0} (\zeta)&=& \mbox{sech}\zeta ,~X_{1} (\zeta)= \zeta \mbox{sech}\zeta.\label{eq23}
\end{eqnarray}
In order to find $T_{0}$ and $T_{1}$,  we
 substitute Eq.(\ref{eq22}) in Eq.(\ref{eq13}) and multiply by $X_0(\zeta)$ and $X_1(\zeta)$ separately,  and after using the orthonormal relations, we get
\begin{subequations}
\begin{eqnarray}  
\qquad\qquad{ T_{1}}_{\tau} (\tau)=\int_{-\infty}^{\infty} F^{(1)} (\zeta,\tau) X_{0} (\zeta)
d\zeta , \label{eq24a}\\
 {T_{0}}_{\tau} (\tau)-2T_{1} (\tau)=-\int_{-\infty}^{\infty} F^{(1)}
 (\zeta,\tau)
X_{1} (\zeta) d\zeta.  \label{eq24b}
\end{eqnarray}
\end{subequations}
As $F^{(1)}(\zeta,\tau)$  given in Eq. (\ref{eq14}) does not contain time $\tau$ explicitly, the right hand side of Eqs. (\ref{eq24a}) and (\ref{eq24b}) are also independent of time,  and hence they give rise to secularities and  the nonsecular conditions can be written as
\begin{subequations}
\begin{eqnarray}
\int_{-\infty}^{\infty}F^{(1)} (\zeta,\tau) X_{0} (\zeta) d\zeta=0, \label{eq25a}\\
\int_{-\infty}^{\infty}F^{(1)} (\zeta,\tau) X_{1} (\zeta) d\zeta=0. \label{eq25b} 
\end{eqnarray}
\end{subequations}
 On substituting the above expressions in Eqs. (\ref{eq24a}) and (\ref{eq24b}), we choose $T_1(\tau)=0$ and  obtain $T_0(\tau)=C$, where $C$  is a constant, which has to be determined. For this, we integrate Eq. (\ref{eq24b}) and obtain 
\begin{eqnarray} 
T_{0}(\tau)&=& \frac{(1+v)}{2}\int_{-\infty}^{\infty}d\zeta ~\zeta F^{(1)}(\zeta,\tau) X_{1}(\zeta)
.\label{eq26}
\end{eqnarray}
\subsection{Variation of soliton parameters}
In order to find the first order correction,   we need to evaluate the eigenstates  explicitly for which we need the values of $m_{t_{1}}$ and $\xi_{t_{1}}$ which can be
 found from the  nonsecularity conditions,  
 by  substituting the
values of $F^{(1)}(\zeta,\tau), X_{0}(\zeta)$ and $X_{1}(\zeta)$ respectively
 from Eqs. (\ref{eq14}) and (\ref{eq23}). The results give the  time evolution of the inverse of the
 width $(m)$  and the velocity $(\xi_{t_{1}})$ of the soliton as
\begin{subequations}
\begin{eqnarray}
m_{t_{1}}&=&-\frac{1}{2v_0} \int_{-\infty}^{\infty} (\beta\left[ {y_{\zeta}\mbox{ sech}\zeta}\right]
_{\zeta}+b\hat\gamma y_{\zeta}\tanh\zeta\mbox{sech}\zeta)\mbox{ sech}\zeta d\zeta, \qquad \label{eq22a}\\
\xi_{t_{1}}&=&-\frac{1}{2m^{2} v_0}\int_{-\infty}^{\infty} (\beta\left[{y_{\zeta}\mbox{ sech}\zeta}\right]
_{\zeta}+b\hat\gamma y_{\zeta}\tanh\zeta\mbox{sech}\zeta) \zeta \mbox{sech}\zeta d\zeta . \label{eq22b}
\end{eqnarray}
\end{subequations}
 In order to evaluate the integrals found in Eqs.~(\ref{eq22a}) and (\ref{eq22b}) explicitly,  we have to substitute the value of `$y$' which we have found by solving  Eq. (\ref{eq5b}).  We consider the most general and meaningful wave solution of Eq. (\ref{eq5b}) suitable for the problem as the periodic function $y=\sin\zeta$.  At this point, it is worth mentioning that Dandoloff and Saxena \cite{ref25} realized that in the case of an XY-coupled spin chain model which is identifiable with our DNA double helical chain model, the ansatz $cn(\zeta, \kappa)$ with the limit $\kappa\rightarrow 0$, energetically favours the periodically deforming spin chain. Hence by substituting $y_{\zeta}=\cos\zeta$ in  Eqs. (\ref{eq22a}) and (\ref{eq22b}) and on evaluating the integrals we obtain
\begin{eqnarray}
m_{t_{1}}=0, \quad
\xi_{t_{1}}=\frac{\pi[\pi \beta-b\hat\gamma(4-\pi)]}{16 m^{2} v_{0}}.\label{eq27}
\end{eqnarray}
 The parameters $m$ and $\xi$ can be written in terms of the original
 variable $\hat t$   as
\begin{eqnarray}
m=m_{0}, \quad
\xi_{\hat {t}}\equiv v=v_{0}+\frac{\epsilon\pi[\beta\pi -b\hat\gamma(4-\pi)]}{16 m^{2} v_{0}},\label{eq28}
\end{eqnarray}
where $1/m_{0}$ is the initial width of the soliton and $v_0$ is the uniform velocity of the soliton in the unperturbed limit. The first of  Eq.(\ref{eq28}) 
says that, the  width $(m^{-1})$ 
of the soliton remains constant. However from the second of Eq. (\ref{eq28}), we
find that the velocity of the 
soliton gets a correction.  It is observed that the correction in  velocity depends on the nature of $\beta$ and $\hat\gamma$
 which can be either positive or negative for $b=\pm 1$. First,  we consider the case corresponding  to $b=1$.
 In this case, when $[\beta\pi-\hat\gamma(4-\pi)]>0$, the velocity of the soliton gets a positive correction and hence soliton may propagate along the DNA chain without forming  a bound state. On the other hand, when $[\beta\pi-\hat\gamma(4-\pi)]\leq 0$, the phonon due to lattice deformation either slows down the soliton or the velocity of the soliton remains unaltered. Finally, if the initial uniform velocity of the soliton before switching on the perturbation due to elastic deformation takes the value  $v_0^2=\frac{\beta\pi-\hat\gamma(4-\pi)}{\beta\pi-\hat\gamma(4-\pi)-16}$, the soliton is stopped by the deformation. The stability of the soliton is guaranteed in all the above cases. A similar argument can be made in the case of $b=-1$ with $[\beta\pi-\hat\gamma(4-\pi)]$  replaced by $[\beta\pi+\hat\gamma(4-\pi)]$. Recently Yakushevich et al \cite{ref36a} and Salerno \cite{refsa} investigated the interaction of soliton with periodic sequence (periodic inhomogeneity),  and the results have very close analogy with our results here. It was shown by them that soliton can easily propagate along DNA without forming a bound state.  It may also be noted that, Zhang et al \cite{ref27} obtained  similar  results in the case of resonant kink impurity interaction and kink scattering in a perturbed sine-Gordon model. In a recent paper, Hwa et al \cite{hwa}, while studying the thermodynamic and dynamic behaviours of twist induced denaturation bubbles in a long, stretched random sequence of DNA using statistical mechanical models, has shown the localization and delocalization of bubbles along the DNA chain. Finally, Eq. (\ref{eq28}) is also similar in form to our recent results of perturbative analysis in the case of an inhomogeneous DNA \cite{ref18,ref26}. Thus, we can say that the lattice deformation gives rise to inhomogeneity in the DNA chain.   
\subsection{First order perturbed soliton}
Now, we explicitly construct the first order correction to the one soliton 
 by substituting  the values of
   $X(\zeta,k),
X_{0}(\zeta), X_{1}(\zeta)$  and $T(\tau,k),~T_{0}(\tau)$ from Eqs.~(\ref{eq21a}), (\ref{eq23}) and (\ref{eq16}),
(\ref{eq26}) and that of $F^{(1)}(\zeta,\tau)$  from Eq.~(\ref{eq14}) and use the
values of $m_{t_{1}}$ and $\xi_{t_{1}}$ from Eqs.~(\ref{eq27}) in Eq.~(\ref{eq22}) to get
\begin{eqnarray}
\Psi^{(1)} (\zeta,\hat t_{0})&= &\frac{1}{\pi}\left[\int_{-\infty}^{\infty}
\frac{dk}{(1+k^{2})^{3}} (1-k^{2}-2ik\tanh\zeta)
  e^{ik\zeta} \right.\nonumber\\
&&\times\int_{-\infty}^{\infty}
d\zeta'(1-k^{2}+2ik\tanh\zeta')
 [\beta\sin\zeta'+\{(\beta-b\hat\gamma)\cos\zeta' \nonumber\\
&&-\frac{\pi}{8}(\pi\beta- b\hat \gamma (4-\pi))\}\tanh\zeta']\mbox{sech}\zeta' e^{-ik\zeta'}\nonumber\\
 &&\times  \left[e^{i\frac{(1+k^{2})}{2k}[\frac{\hat t_0}{m}-(1+v)(\zeta-\zeta')]}-1\right]
 +(1+v)\mbox{sech}\zeta \nonumber\\
&&\times \int_{-\infty}^{\infty}
d\zeta' ~\zeta^{'2}[\beta\sin\zeta'+\{(\beta-b\hat\gamma )\cos\zeta'-\frac{\pi}{8}(\pi\beta- b\hat \gamma (4-\pi))\}\nonumber\\
&&\left.\times\tanh\zeta']
 \mbox{sech}^{2}\zeta'\right] .\label{eq29}
\end{eqnarray}
We evaluate the integrals in Eq. (\ref{eq29})  by finding the values of the residues at poles of different orders using residue theorem (for details see \cite{ref18,ref24,ref28}). After lengthy algebra and some approximations  the explicit form  of the perturbed  kink (upper sign)-antikink (lower sign) one soliton solution  in terms of the original variables is written  as
 \begin{eqnarray}
 \Psi(z,t_{0})&\approx &4~
\mbox{arc}\tan\exp[m_{0}(z-v_{0} t_{0})]\nonumber\\
&&+\frac{\epsilon\pi[\pi \beta-b\hat\gamma(4-\pi)]}{16 m^{2} v_{0}}
 \left[m(v^2-1)+2 v t_{0}\right]\mbox{sech}[m(z-v t_{0})].\label{eq30}
 \end{eqnarray}
Having found $\Psi(z,t_0)$ we find  $\phi(z,t_{0})$ using the relation $\phi=\frac{\Psi}{2}
$  and  plot the same   in Figs. 3(a,b)    by choosing $\beta=\hat\gamma =1,b=-1$ and  $v_{0}=0.4$. From the figures,  we observe that the 
lattice deformation introduces only small fluctuations   in the form of periodic oscillations closely resembling the shape of the lattice deformation in the width of the soliton  (see Figs. 3(a,b)). We have schematically represented this in Fig. 3(c), where the dotted line along the strands (lattice) represent the periodic deformation of the lattice. It shows that 
the  lattice deformation  in DNA does not affect  opening of bases
in DNA double helix.

\section{Conclusion }
 In this paper, we  studied the effect of   phonon  interaction on base pair opening in DNA  by considering the 
 dynamic plane base rotator
 model. The dynamics of this model in the continuum limit gives rise to a perturbed sine-Gordon equation coupled with a linear wave equation representing  longitudinal lattice vibration, which were derived from the Hamiltonian consisting of the stacking energy, hydrogen bonding energy, energy corresponding to the lattice deformation and its coupling with the stacking and hydrogen bonding energy. In the unperturbed limit, the dynamics is governed by the kink-antikink soliton of the integrable sine-Gordon equation which represents the opening of base pairs in  DNA without lattice deformation.  In order to understand the effect of lattice deformation on the base pair opening, we carried out a perturbation analysis using multiple-scale soliton perturbation theory. From the results of variation of soliton parameters we observe that when the DNA lattice  deforms in a periodic way, the width of the soliton remains constant. However,  the  velocity of the soliton  increases or decreases or remains uniform or even the soliton stops depending on the values of the coupling strengths $\beta$ and $\hat\gamma$.  Interestingly, the soliton in all the  above cases are found to be stable. From the results of the perturbed soliton we observe that the periodic lattice deformation introduces fluctuation in the width of the soliton. However, there is no change in the topological character of the soliton in the asymptotic region. The above dynamical behaviour may act as energetic activators of the enzyme transport during the process of transcription in DNA.    
\section*{Acknowledgements}
   The work of M. D and V.V   forms  part of a major  DST  project. 
  
\newpage
Fig. 1. (a) A schematic representation of the structure of  B-form   DNA double helix. (b) A horizontal projection of the $n^{th}$ base
pair in the xy-plane.\\
Fig. 2. (a) Kink  and (b) antikink  one soliton solutions of the sine-Gordon equation (Eq. (\ref{eq5a}) when $\epsilon=0$). (c) A sketch of the formation of open state configuration in terms of
kink-antikink solitons in a DNA double helical chain.\\
Fig. 3.  The perturbed (a) kink-soliton and (b)  antikink-soliton with  $\beta=\hat\gamma = 1.0, b=-1.0$ and  $v_0=0.4$. (c) A  sketch of the open state configuration in DNA  
 with small fluctuations and periodic deformation in the lattice.\\
\newpage
 \begin{figure}
\begin{center}
\psfig{file=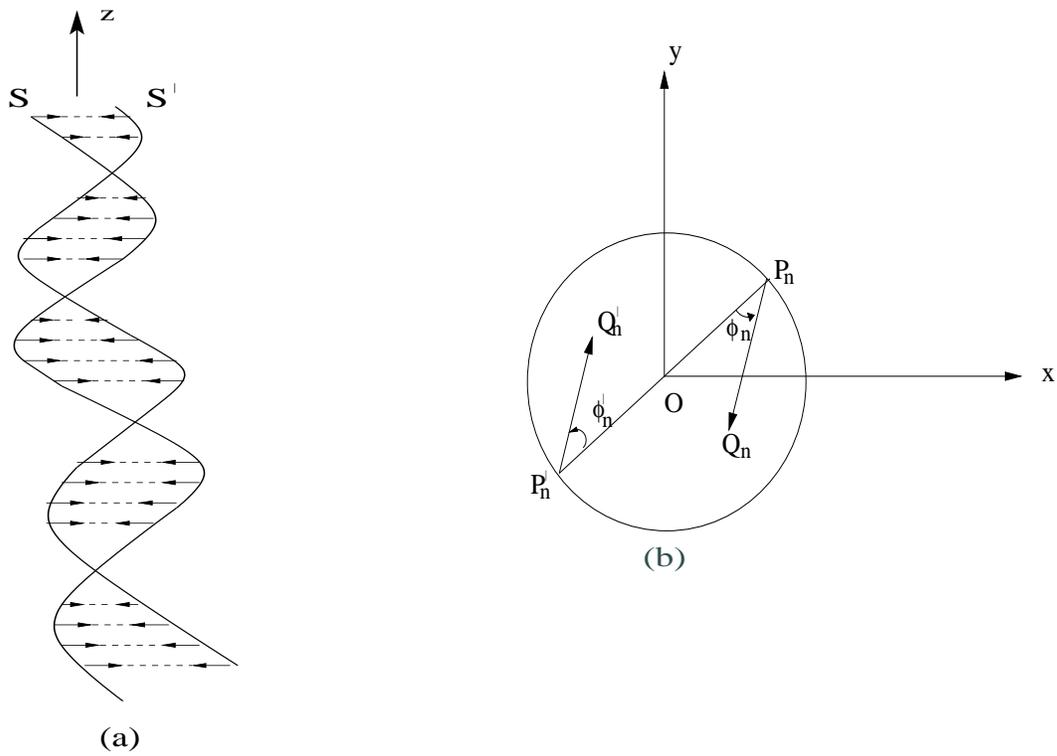,height =10cm, width=14cm}
\caption{(a) A schematic representation of the structure of  B-form   DNA double helix. (b) A horizontal projection of the $n^{th}$ base
pair in the xy-plane.}
\end{center}
\end{figure}
\begin{figure}
\begin{center}
\epsfig{file=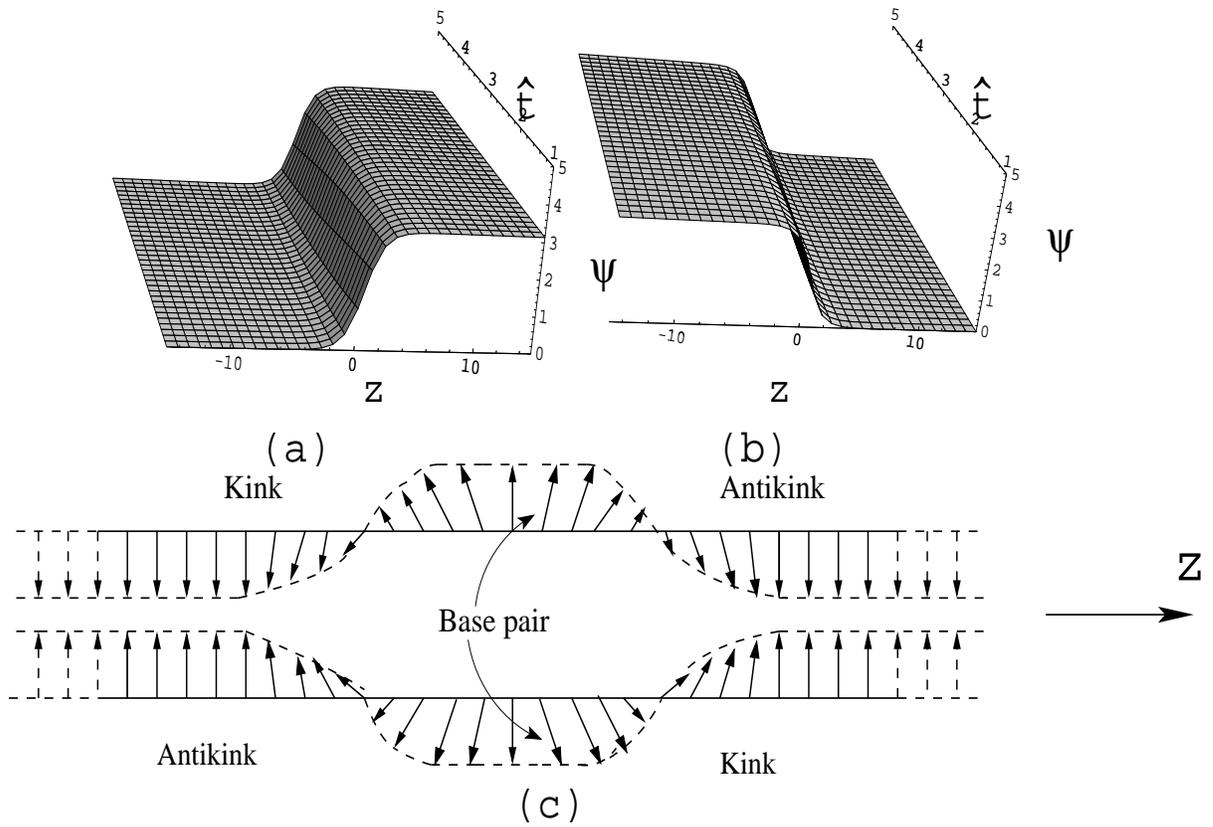,height =14cm, width=16cm}
\caption{(a) Kink  and (b) antikink  one soliton solutions of the sine-Gordon equation (Eq. (\ref{eq5a}) when $\epsilon=0$). (c) A sketch of the formation of open state configuration in terms of
kink-antikink solitons in a DNA double helical chain.}
\end{center}
\end{figure}
 \begin{figure}
\begin{center}
\epsfig{file=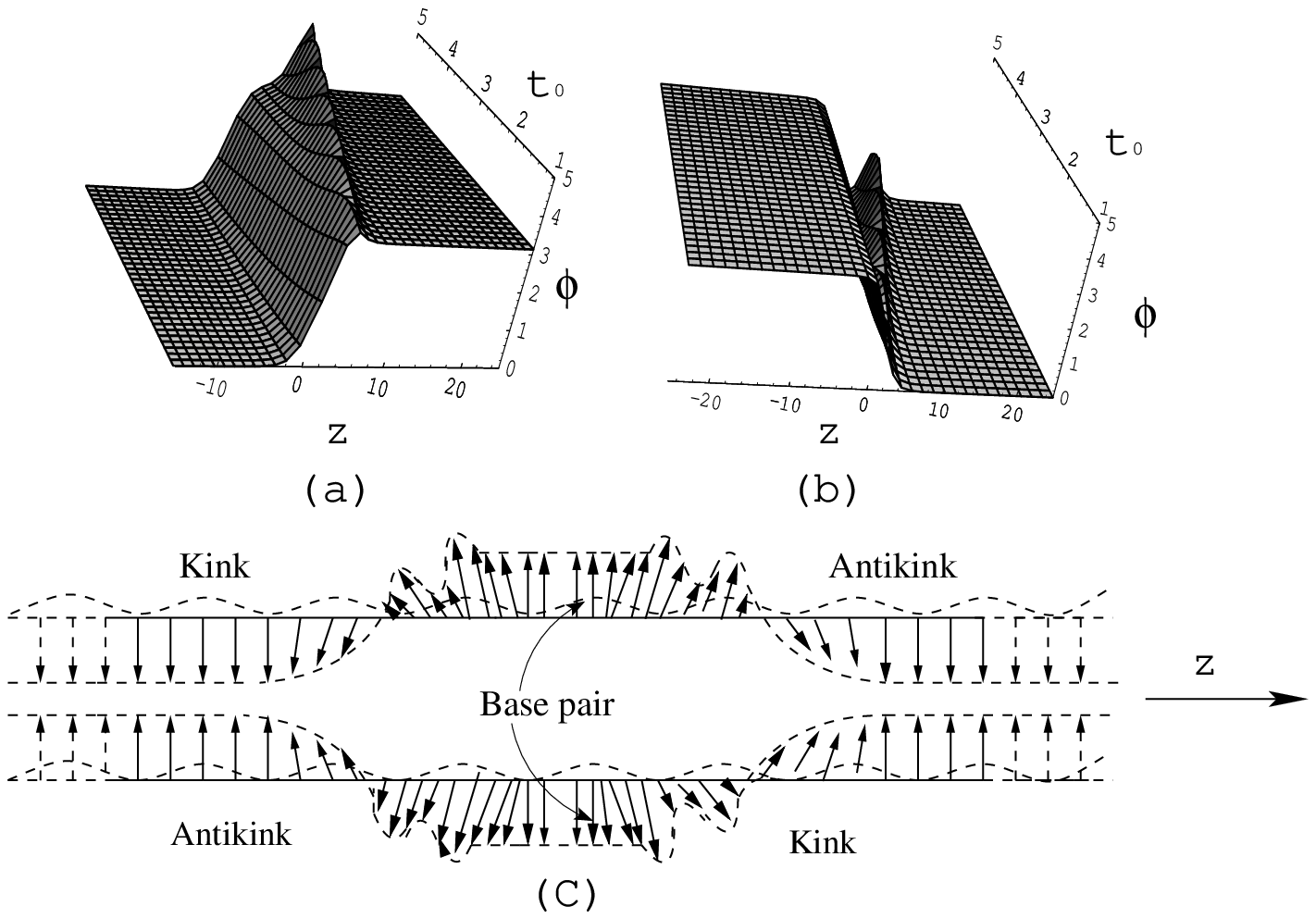,height =12cm, width=16cm}
\caption{ The perturbed (a) kink-soliton and (b)  antikink-soliton with  $\beta=\hat\gamma = 1.0,  b=-1.0$ and  $v_0=0.4$. (c) A  sketch of the open state configuration in DNA  
 with small fluctuations and periodic deformation in the lattice }
\end{center}
\end{figure}

\begin{thebibliography}{03}
\bibitem{ref1}
L. V. Yakushevich, { \it Nanobiology} {\bf 1} (1992) 343.
\bibitem{ref2}
S. W. Englander, N. R. Kallenbanch, A. J. Heeger, J. A. Krumhansl and S. Litwin,
  Proc. Natl. Acad. Sci. USA  {\bf77} (1980) 7222.
\bibitem{ref11}
S. Yomosa, Phys. Rev. A {\bf 27}  (1983) 2120.
\bibitem{ref12}
S. Yomosa, Phys. Rev. A {\bf 30} (1984) 474.
\bibitem{ref13}
S. Takeno and S. Homma,  Prog. Theor. Phys. {\bf 70} (1983) 308.
\bibitem{ref14}
S. Takeno and S. Homma,  Prog. Theor. Phys. {\bf 72} (1984) 679.
\bibitem{ref3}
M. Peyrard and A. R. Bishop,  Phys. Rev. Lett. {\bf 62} (1989) 2755.
\bibitem{ref4}
P. L. Christiansen, P. C. Lomdahl and V. Muto,  Nonlinearity {\bf 4} (1990) 477.
\bibitem{refpb1}
T. Dauxios, Phys. Lett. A {\bf 159}   (1991) 390.
\bibitem{ref14a}
S. Takeno, Phys. Lett. A {\bf 339} (2005) 352. 
 \bibitem{refsa}
M. Salerno, Phys. Rev. A {\bf 44}  (1991) 5292.
\bibitem{g1}
G. Gaeta,   Phys. Lett. A {\bf 143}  (1990) 227.
\bibitem{g2}
G. Gaeta,   Phys. Lett. A {\bf 168} (1992) 383.
\bibitem{ref36a}
L. V. Yakushevich, A. V. Savin and L. I. Manevitch, Phys. Rev. E {\bf 66} (2002) 016614.
\bibitem{g3}
G. Gaeta, Phys. Rev. E {\bf 74} (2006) 021921.
\bibitem{g4}
M. Cadoni, R. De Leo and G. Gaeta,    Phys. Rev. E {\bf 75} (2007) 021919.
\bibitem{ref18}
M. Daniel and V. Vasumathi, Physica D {\bf 231} (2007) 10.
\bibitem{ref26}
M. Daniel and V.Vasumathi, Phys. Lett. A {\bf 372}  (2008) 5144.
\bibitem{ref15}
C. R. Calladine, J. Mol. Biol. {\bf 161}  (1982) 343.
\bibitem{ref16}
A. C. Scott, Phys. Lett. A {\bf 86}  (1981) 60.
\bibitem{ref16a}
A. S. Davydov, {\it Solitons in Molecular Systems} (Reidel, Dordrecht, 1985) PP. 1-20.
\bibitem{refdvs1}
A. S. Davydov, Ukr. Fiz. Zh. {\bf 20}  (1975) 179.
\bibitem{ref17}
J. X. Xiao, J. T. Lin and G. X. Zhang, J. Phys. A. Math. Gen. {\bf 20} (1987) 2425.
\bibitem{pople}
A. Pople, Proc. R. Soc. London  Ser. A {\bf 205} (1951) 165.
\bibitem{ref19}
M. J. Ablowitz, D. J. Kaup, A. C. Newell, and H. Segur, Stud. Appl. Math. {\bf 53}  (1974) 249.
\bibitem{ref20}
D. J. Kaup, SIAM  J. Appl. Math. {\bf 31}  (1976) 12.
\bibitem{ref21}
G. L. Lamb, {\it Elements of Soliton Theory} (Wiely, New York, 1980) PP. 259-278.
\bibitem{ref22}
J. Yan and Y. Tang, Phys. Rev. E {\bf 54} (1996) 6816.
\bibitem{ref23}
Y. Tang and W. Wang, Phys. Rev. E {\bf 62}  (2000) 8842.
\bibitem{ref24}
J. Yan, Y. Tang and G.  Zhou, Phys. Rev. E {\bf 58}  (1998) 1064.
\bibitem{ref25}
R. Dandoloff and A. Saxena, J. Phys.: Condens. Matter {\bf 9} (1997) L667.
\bibitem{ref27}
F. Zhang, Y. S. Kivshar and L. Vazquez, Phys. Rev. A  {\bf 45}  (1992)  6019.
\bibitem{hwa}
T. Hwa, E. Marinari, K. Sneppen and L. Tang, PNAS {\bf 100} (2003) 4411.
\bibitem{ref28}
 E. Kreyszig, {\it  Advanced Engineering  Mathematics} (John-Wiley, New York, 2002) PP. 771-794. 
\end{thebibliography}
\end{document}